\documentclass[prb,twocolumn,showpacs]{revtex4-1}
\usepackage{amsmath}
\usepackage{amsfonts}
\usepackage{graphicx}
\newcommand{\beq}{\begin{equation}}
\newcommand{\eeq}{\end{equation}}
\newcommand{\be}{\begin{equation}}
\newcommand{\ee}{\end{equation}}
\newcommand{\ra}{\right\rangle}
\newcommand{\la}{\left\langle}

\begin{document}

\title{Statistics of quantum transport in weakly non-ideal chaotic cavities}

\author{Sergio Rodr\'iguez-P\'erez$^1$, Ricardo Marino$^2$,
Marcel Novaes$^1$ and Pierpaolo Vivo$^2$}

\affiliation{1) Departamento de F\'isica, Universidade Federal de S\~ao Carlos, S\~ao
Carlos, SP, 13565-905, Brazil}\affiliation{2) Laboratoire de Physique Th\'eorique et
Mod\`eles Statistiques, UMR CNRS 8626, Universit\'e Paris-Sud, 91405 Orsay, France}

\begin{abstract}
We consider statistics of electronic transport in chaotic cavities where time-reversal
symmetry is broken and one of the leads is weakly non-ideal, i.e. it contains tunnel
barriers characterized by tunneling probabilities $\Gamma_i$. Using symmetric function
expansions and a generalized Selberg integral, we develop a systematic perturbation
theory in $1-\Gamma_i$ valid for arbitrary number of channels, and obtain explicit
formulas up to second order for the average and variance of the conductance, and for the
average shot-noise. Higher moments of the conductance are considered to leading order.
\end{abstract}

\maketitle

\section{Introduction}
At low temperatures and applied voltage, provided that the average electron dwell time is
well in excess of the Ehrenfest time, statistical properties of electronic transport in
mesoscopic cavities exhibiting chaotic classical dynamics are universal. Random matrix
theory (RMT) has proven very successful in describing this universality
\cite{beenakker_transport}. In this approach, the scattering $S$-matrix of the cavity is
modeled by a random unitary matrix\cite{baranger_mello_94,jalabert_pichard_beenakker_94}
(see Refs. \onlinecite{scatt1,scatt2,scatt3} for the most recent analytical results on
distribution of $S$).

We consider a chaotic cavity attached to two leads, having $N_1$ and $N_2$ (with $N_1\leq
N_2$) open channels in each lead, and denote by $N=N_1+N_2$ the total number of channels.
The $S$ matrix can be written in the usual block form $S= \left(\begin{matrix} r &
t^\prime
\\ t & r^\prime\end{matrix} \right), $ in terms of reflection and transmission matrices.
Landauer-B\"uttiker theory \cite{landauer_57,fisher_lee_81,Imry,butt1,butt2} expresses
most physical observables in terms of the eigenvalues $\{T_1,\ldots,T_{N_1}\}$ of the
hermitian matrix $t t^\dagger$. A figure of merit is the conductance
$G(T)=G_0\sum_{i=i}^{N_1}T_i$, where $G_0 = 2e^2/h$ is the conductance quantum. The
assumption that $S$ is a random matrix implies that the $T_j$'s are correlated random
variables characterized by a certain joint probability density (jpd), and as a
consequence every observable becomes a random variable whose statistics is of paramount
interest.

When the leads attached to the cavity are ideal, $S$ is uniformly distributed in one of
Dyson's circular ensembles of random matrices, which are labeled by a parameter $\beta$:
it is unitary and symmetric for $\beta=1$ (corresponding to systems that are invariant
under time-reversal), just unitary for $\beta=2$ (broken time-reversal invariance) and
unitary self-dual for $\beta=4$ (anti-unitary time-reversal invariance). In this ideal
case the jpd of reflection eigenvalues $R_i=1-T_i$ is
given\cite{baranger_mello_94,jalabert1,forrester} by the Jacobi ensemble of RMT, namely
\be\label{jpdideal} P_\beta^{(0)}(R)\propto
|\Delta(R)|^\beta\prod_{i=1}^{N_1}(1-R_i)^{\frac{\beta}{2}(N_2-N_1+1)-1},\ee where
$\Delta(R)=\prod_{j<k}(R_k-R_j)$ is the Vandermonde determinant. The average and variance
of the conductance were studied, using perturbation theory in $1/N$, long
ago.\cite{baranger_mello_94, beenakker_93, iida_weidenmuller_90,
jalabert_pichard_beenakker_94} In particular, as $N\to\infty$ this variance becomes a
constant depending only on the symmetry class, a phenomenon that has been dubbed
\emph{universal conductance fluctuations}. More recently, a fruitful approach based on
the theory of the Selberg integral was developed \cite{sommers_savin_06} and afterwards
extended \cite{sommers_wieczorek_savin_07, savin_sommers_wieczorek_08,novaes_08,
sommers_savin_schur} to compute transport statistics non-perturbatively. The full
distribution of $G$, known to be strongly non-Gaussian for a small number of channels
\cite{baranger_mello_99, sommers_savin_schur, kumar_pandey_10,
sommers_wieczorek_savin_07}, was studied in \cite{kanzieper_ozipov_08,
kanzieper_ozipov_09,vivo_majumdar_bohigas_08, vivo_majumdar_bohigas_10}. The statistics
of other observables was studied in \cite{novaes_07, novaes_08, vivo_vivo_08,
livan_vivo_11,texier}. The integrable theory of quantum transport in the ideal case,
pioneered in\cite{kanzieper_ozipov_08, kanzieper_ozipov_09} for $\beta=2$, has been
recently completed including the other symmetry classes.\cite{mezzadri_simm_11}

A more generic situation occurs when the leads are not ideal but contain tunnel barriers.
A simple barrier, which does not mix transversal modes, is represented by a set of
tunneling probabilities $\{\Gamma_i\}$, one for each open channel. In this case the
distribution of $S$ is given by the so-called Poisson kernel
\cite{baranger_mello_99,brouwer_95} \be \mathcal{P}_\beta(S)=
\left[\det(1-\bar{S}S^\dagger)\det(1-S\bar{S}^\dagger)\right]^{\beta/2 -1 -\beta
N/2},\label{poissonkernel} \ee which depends only on $\beta$ and the average scattering
matrix $\bar{S}$ (whose singular values are determined by the tunneling probabilities of
each lead). In the limit $\Gamma_i\to 1$ we have $\bar{S}=0$ and one recovers the ideal
case. Even though controllable barriers in the leads are by now an established
experimental protocol\cite{gustav}, few explicit theoretical predictions are available
due to the complicated nature of the Poisson kernel. For instance, the average and
variance of conductance are only known perturbatively in the limit $\Gamma_i N\gg 1$ for
all $i$\cite{iida_weidenmuller_90,efetov,BB}. Semiclassical studies of transport in the
large $N$ limit and non-ideal setting have also recently
appeared.\cite{semic0,semic,semic2}

A more systematic RMT theoretical investigation was initiated when Vidal and Kanzieper
\cite{kanzieper_vidal} obtained the jpd of reflection eigenvalues for $\beta=2$ and only
one non-ideal lead. In this work we characterize those $N_1$ non-ideal channels by a
diagonal matrix $\gamma=\mathrm{diag}(\gamma_1,\ldots,\gamma_{N_1})$, with
$\gamma_i=1-\Gamma_i$ (these are not the same $\gamma_i$ which appear in
Ref.\onlinecite{kanzieper_vidal}; the definitions differ by a square root). The other
lead is kept ideal. Our goal is to use symmetric function expansions and a generalized
Selberg integral to develop a systematic perturbation theory in $\gamma$ of this problem.
In this framework we present explicit formulas for the most useful transport statistics.
In contrast to most previous results, ours are valid for an arbitrary number of channels
in the two leads, i.e. they are not restricted to the large-$N$ limit.

\section{Perturbative $\gamma$-expansion}

Let $(a)_n=a(a+1)\cdots (a+n-1)$ be the rising factorial and let \be _2
F_1(a,b;c;x)=\sum_{n\ge 0} \frac{(a)_n(b)_n}{(c)_nn!}x^n\ee be the hypergeometric
function. Let $\mathcal{F}$ be the $N_1\times N_1$ matrix whose elements are \be
\mathcal{F}_{ij}={}_2 F_1(N_2+1,N_2+1;1;\gamma_iR_j).\ee When the non-ideal lead supports
$N_1$ channels, the jpd of reflection eigenvalues is given by \cite{kanzieper_vidal} \be
P_2^{( \gamma)}(R)=Z\ {\det}^{N} (1-\gamma)
\frac{\Delta(R)}{\Delta(\gamma)}{\det}(\mathcal{F})\prod_{i=1}^{N_1}(1-R_i)^{N_2-N_1},\label{kanzr}\ee
where $Z$ is a normalization constant, \be Z=\frac{N!}{N_1!N_2!}\prod_{i =
1}^{N_1}\frac{(N_2)!^2}{(N_2+i)!(N_2-i)!}. \ee The expression \eqref{kanzr} is hardly
operational. We therefore start by writing it in a perturbative way, i.e. as an infinite
series in $\gamma$.

Let a non-increasing sequence of positive integers $\lambda_1,\lambda_2,\ldots$ be called
a partition of $n$ if $\sum_i\lambda_i=n$ and let this be denoted by $\lambda\vdash n$.
The number of parts in $\lambda$ is $\ell(\lambda)$ and we assume $\lambda_m=0$ if
$m>\ell(\lambda)$. Partitions can be used to label a very important set of symmetric
polynomials known as Schur polynomials, which are denoted by $s_\lambda$. Assuming $N_1$
variables, they are defined by \be
s_\lambda(x)=\frac{1}{\Delta(x)}\det\left(x_j^{\lambda_i-i+N_1}\right).\ee For example,
the first few such polynomials are given by
\begin{align} s_0(x)=1&, \quad s_1(x)=\sum_{i=1}^{N_1}x_i,\\ s_{11}(x)=\sum_{i<j}^{N_1}x_ix_j&,
\quad s_2(x)=s_{11}(x)+\sum_{i=1}^{N_1}x_i^2.\end{align} If we define \be
\alpha_{\lambda}=\prod_{i=1}^{N_1}\binom{N+\lambda_i-i}{N_2}^2,\ee the following
expansion can be established: \be \mathrm{det}(\mathcal{F})=
\Delta(\gamma)\Delta(R)\sum\limits_{\lambda}\alpha_{\lambda}s_\lambda(\gamma)s_\lambda(R),
\label{expansioneq}\ee where the infinite sum is over all possible partitions. This
follows from the nice structure of $\mathcal{F}_{ij}$, which depends on the indices $ij$
only through the combination $\gamma_iR_j$. An account of this and similar identities can
be found for example in the book by Hua\cite{hua}.

In order to use \eqref{expansioneq} to express the jpd of reflection eigenvalues, it is
useful to factor out the $\alpha_0$ term and notice that \be
\frac{\alpha_\lambda}{\alpha_0}=\frac{[N]^2_\lambda}{[N_1]^2_\lambda},\label{factoralpha}\ee
where \be [N]_\lambda=\prod_{i=1}^{\ell(\lambda)}\frac{(N+\lambda_i-i)!}{(N-i)!}\ee is a
generalization of the rising factorial. The normalization constant then simplifies as \be
Z'=Z\ \alpha_0=\prod_{i=1}^{N_1}\frac{(N-i)!}{(N_1-i)!(N_2-i)!i!}.\ee This is precisely
the normalization constant missing from (\ref{jpdideal}). Finally, combining
\eqref{jpdideal}, \eqref{kanzr}, \eqref{expansioneq} and \eqref{factoralpha} we get the
final result,\be\frac{P_2^{(\gamma)}(R)}{P_2^{(0)}(R)}=Z'{\rm det}^{N}
(1-\gamma)\sum\limits_{\lambda}\frac{[N]^2_\lambda}{[N_1]^2_\lambda}
s_\lambda(\gamma)s_\lambda(R).\label{jpdnonideal}\ee

\section{Computing observables}

Since any observable is a symmetric function of the reflection eigenvalues, it must be
expressible as a linear combination of Schur polynomials; hence it suffices to obtain the
average value of $s_\mu(R)$ for an arbitrary partition $\mu$. In this way we are led to
consider the multiple integral \be \int_0^1
\Delta^2(R)s_\lambda(R)s_\mu(R)\prod_{i=1}^{N_1}(1-R_i)^{N_2-N_1}dR,\ee (where $dR\equiv
\prod_j dR_j$) which is a generalization of Selberg's integral\cite{importance}. However,
this is difficult to evaluate directly. One way to proceed is to express the product of
two Schur polynomials again as a linear combination of Schur polynomials, \be
s_\lambda(R)s_\mu(R)=\sum_{\nu}C_{\lambda\mu}^\nu s_\nu(R),\ee where the constants
$C_{\lambda\mu}^\nu$ are known as Littlewood-Richardson coefficients.\cite{sagan} There
is no explicit formula for them, but they can be computed using some recursive algorithms
and there are tables for the first ones. For instance, the coefficients with $\nu$ up to
$4$ are given by


\begin{align*}
s_0s_\lambda=s_\lambda, &\quad s_1s_1=s_2+s_{11},\\
s_2s_1=s_3+s_{21}, &\quad s_{11}s_1=s_{111}+s_{21} , \\
s_3s_1=s_4+s_{31},&\quad
s_{21}s_1=s_{31}+s_{22}+s_{211},\\
 s_{2}s_2=s_4+s_{31}+s_{22},&\quad
s_2s_{11}=s_{31}+s_{211},\\
s_{111}s_1=s_{1111}+s_{211}, &\quad
s_{11}s_{11}=s_{1111}+s_{211}+s_{22}.
\end{align*}

By means of the Littlewood-Richardson coefficients, we only need to consider the simpler
integral \be \mathcal{I}_\nu=\int_0^1
\Delta^2(R)s_\nu(R)\prod_{i=1}^{N_1}(1-R_i)^{N_2-N_1}dR, \ee which is known to be given
by\cite{kaneko,kadell} \be
\mathcal{I}_\nu=s_\nu(1^{N_1})\prod_{i=1}^{N_1}\frac{i!(N_1+\nu_i-i)!(N_2-i)!}{(N+\nu_i-i)!},\ee
where $s_\nu(1^{N_1})$ is the value of a Schur polynomial when all its arguments are
equal to unity.

If we combine the above result with $Z'$ we get a substantial simplification, which is
manifestly a rational function of $N_1$ and $N_2$, i.e. the variable $N_1$ no longer
appears as the limit to products. The result is \be Z'\mathcal{I}_\nu=\frac{[N_1]_\nu^2
\chi_\nu(1)}{[N]_{\nu}|\nu|!},\ee where $\nu\vdash|\nu|$ and $\chi$ is the character
function in the permutation group, so $\chi_\nu(1)$ is the dimension of the irreducible
representation of that group associated with partition $\nu$ (to arrive at this result we
have used that $s_\nu(1^{N_1})=\chi_\nu(1)[N_1]_\nu/n!$).

The final result is that the average value of $s_\mu(R)$, with respect to the jpd
(\ref{jpdnonideal}), is given by \be \langle s_\mu(R)\rangle_\gamma={\rm det}^{N}
(1-\gamma)\sum\limits_{\lambda}D_{\mu\lambda}s_\lambda(\gamma),\ee with \be
D_{\mu\lambda}=\frac{[N]_\lambda^2}{[N_1]_{\lambda}^2}\sum_\nu C_{\lambda,\mu}^\nu
\frac{[N_1]_\nu^2 \chi_\nu(1)}{[N]_{\nu}|\nu|!}.\ee

\section{The leading order}

The jpd (\ref{jpdnonideal}) equals the jpd of the ideal case \eqref{jpdideal} times a
correction which can be systematically expanded in powers of $\gamma$. In this way any
observable in the finite-$\gamma$ regime can be expressed in terms of observables
computed in the ideal regime. For example, to leading order we have \be
\frac{P_2^{(\gamma)}(R)}{P_2^{(0)}(R)}\propto
\left[1+\frac{N}{N_1}\left(\frac{N}{N_1}s_1(R)-N_1\right){\rm
Tr}\gamma\right].\label{expjpd}\ee

As a first application, let $\la G^n \ra_{\gamma}$ be the average value of the $n$th
moment of the conductance in the non-ideal case. Using \eqref{expjpd} and the fact that
$s_1(R)=N_1-G$, it is easy to see that the difference between the weakly non-ideal case
and the ideal case is given to leading order by \be\la G^n \ra_{\gamma}-\la G^n
\ra_{0}\approx\frac{N}{N_1}{\rm Tr}\gamma
 \left[N_2\la G^n \ra_{0} - \frac{N}{N_1}\la G^{n+1} \ra_{0}\right].\label{main_formula}
\ee A similar estimate holds for other transport statistics.

\section{Statistics of conductance up to second order}

Using the approach presented here, the average value of any observable can in principle
be found to any order in $\gamma$. Consider for instance the average conductance. In the
ideal case, it is given by $ \la G \ra_{0} = N_1N_2/N$. Up to second order in $\gamma$,
the calculations we just outlined provide \be \la G \ra_{\gamma}\approx\frac{N_1 N_2}{N}
- \frac{N_2^2\mathrm{Tr}\gamma}{N^2-1}+\frac{N_2^2\,[2\mathrm{Tr}\gamma^2-
N(\mathrm{Tr}\gamma)^2]}{(N^2-1)(N^2-4)}.\label{Gaverage}\ee

We stress that this result is exact for any number of channels $N_1$ and $N_2$. It is
interesting to check that it is compatible with the available results in the
literature,\cite{BB} which are on the contrary exact in $\gamma$ but perturbative in
$N_1,N_2$. Taking into account that traces are of order $N_1$, we see that each term in
(\ref{Gaverage}) scales linearly with $N$. For example, when $N_1=N_2$, we have \be
\lim_{N_1=N_2\to\infty}\frac{\la G\ra_\gamma}{N_1} \approx\frac{1}{2}-\frac{{\rm
tr}\gamma}{4}-\frac{({\rm tr}\gamma)^2}{8},\label{limitinglaw}\ee where ${\rm tr}$ is the
normalized trace, \be {\rm tr}X=\lim_{N_1\to\infty}\frac{1}{N_1}{\rm Tr}X.\ee The
limiting law \eqref{limitinglaw} perfectly matches the result by Brouwer and Beenakker
[eq. 6.17 in Ref.\onlinecite{BB}] which reads
\begin{equation}
\langle G\rangle_\gamma^{(BB)} \approx \frac{g_1
g_1^\prime}{g_1+g_1^\prime}+\mathcal{O}(N^{-1}),
\end{equation}
where in our notation $g_1=N_1-\mathrm{Tr}\gamma$ and $g_1^\prime=N_2$. Computing the
same limit as in \eqref{limitinglaw} we get \be \lim_{N_1=N_2\to\infty}\frac{\la
G\ra_\gamma^{(BB)}}{N_1}
\approx\frac{1-\mathrm{tr}\gamma}{2-\mathrm{tr}\gamma},\label{limitinglaw2}\ee whose
expansion in $\gamma$ up to the second order precisely reproduces \eqref{limitinglaw}.
Notice that the average conductance decreases with $\gamma$, as should be expected.

\begin{figure}[!t]
\includegraphics[scale=0.4,clip]{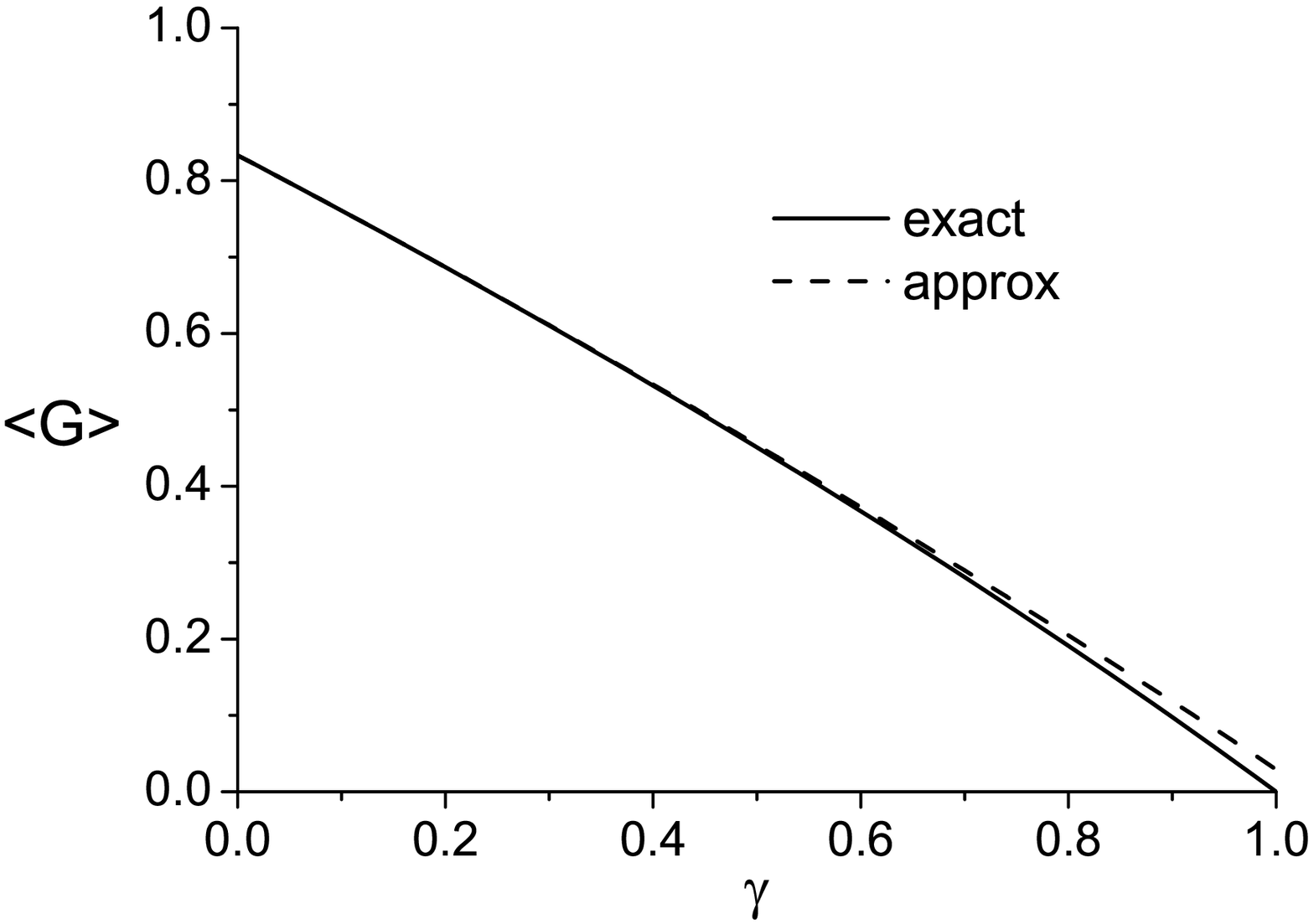}
\includegraphics[scale=0.42,clip]{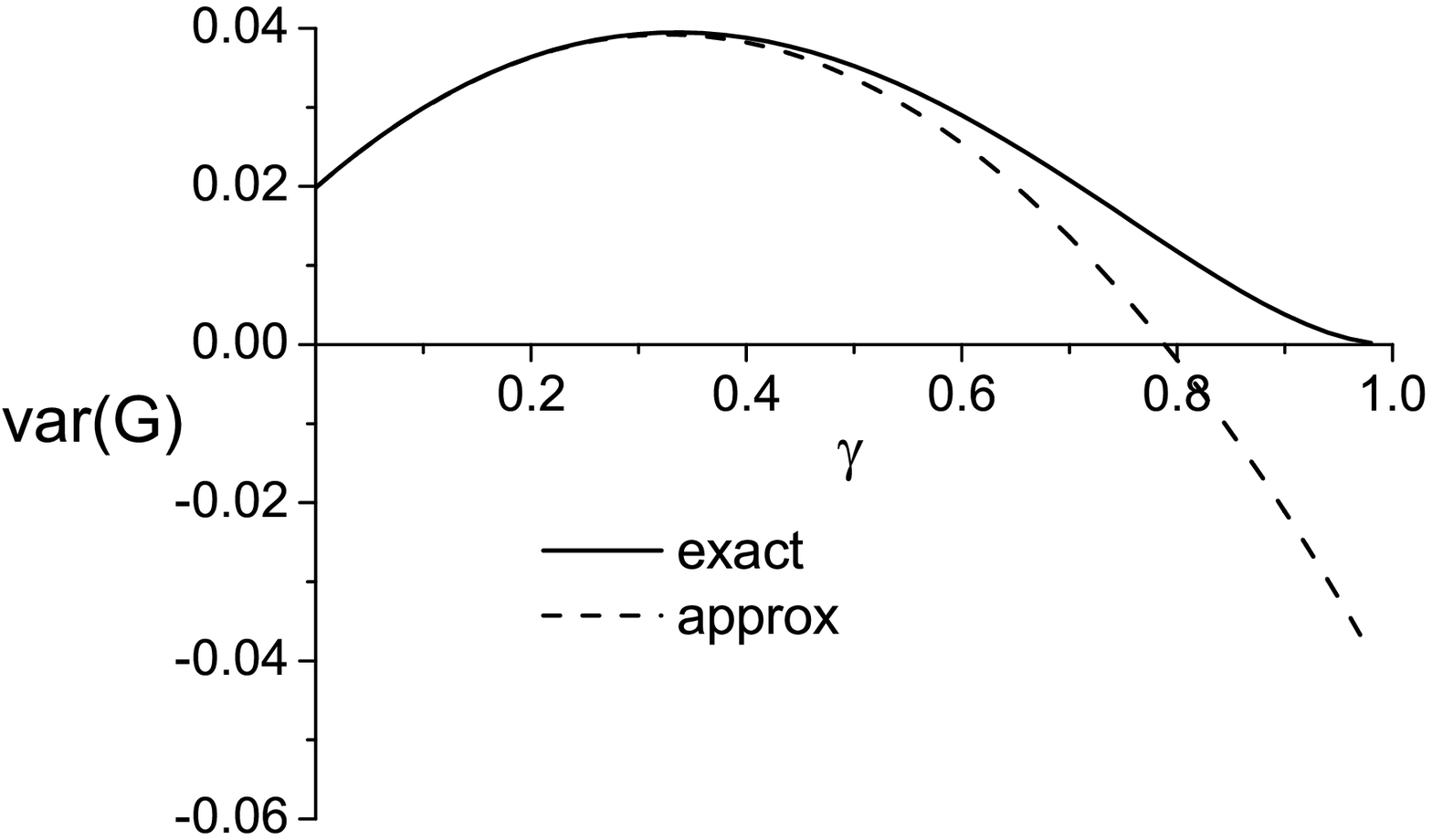}
\caption{Average and variance of conductance as functions of $\gamma$ for $N_1=1$ and
$N_2=5$. Solid and dashed lines are, respectively, exact results and our approximations.
For the average, the difference is minimal. For the variance, the approximation predicts
a non-physical negative result at high $\gamma$, but is excellent up to moderate values
of $\gamma$.}
\end{figure}

The variance of conductance, on the other hand, is given in the ideal case by \be {\rm
var}_0G=\frac{N_1^2N_2^2}{N^2(N^2-1)}.\ee In the non-ideal case, up to second order in
$\gamma$, it becomes \begin{align} \mathrm{var}_\gamma(G)\approx\frac{N_1^2
N_2^2}{N^2(N^2-1)} + \frac{2\,
N_2^2(N_1-N_2)^2\mathrm{Tr}\gamma}{N(N^2-1)(N^2-4)}+\nonumber \\\frac{N_2^2[A_1N
(\mathrm{Tr}\gamma)^2+2B_1(N^2-1)\mathrm{Tr}\gamma^2]}{N(N^2-1)^2(N^2-4)(N^2-9)}&\label{varapprox},
\end{align}
where \be A_1=(N_1-2N_2)(3N_1-8N_2)N^2+ 20 N_1 N_2-37N_2^2-3,\ee and \be
B_1=(N_1-2N_2)(N_2N^2+N_2-5N_1).\ee Again, the result is exact for any $N_1,N_2$. Notice
that each order in $\gamma$ attains a finite value in the large $N$ limit.  For instance,
when $N_1=N_2$, the first order vanishes identically and we have \be
\lim_{N_1=N_2\to\infty}\mathrm{var}_\gamma(G)\approx\frac{1}{16}+\frac{5({\rm
tr}\gamma)^2-4{\rm tr}\gamma^2}{64}.\ee which perfectly matches the first terms of the
expansion in $\gamma$ of formula 6.24 in Ref.\onlinecite{BB}, where in our notation
$g_1=N_1-\mathrm{Tr}\gamma$, $g_2=N_1-2 \mathrm{Tr}\gamma+\mathrm{Tr}\gamma^2$,
$g_3=N_1-3 \mathrm{Tr}\gamma+3\mathrm{Tr}\gamma^2-\mathrm{Tr}\gamma^3$ and
$g_p^\prime=N_2$ for all $p$.

As a further check, we consider the case $N_1=1$, for which the full density of
conductance is known\cite{kanzieper_vidal} in terms of a single scalar opacity parameter
$\gamma$,
\begin{equation}
f_\gamma(G)=N_2 G^{N_2-1}\phi_\gamma(G),\label{formulaG}
\end{equation}
where
\begin{equation}
\phi_\gamma(G)=(1-\gamma)^{N_2+1}\ _2 F_1(N_2+1,N_2+1;1;\gamma(1-G)).
\end{equation}
The average conductance (and similarly for the variance) is given by the integral
$\langle G\rangle_\gamma =\int_0^1 dG\ G f_\gamma(G)$. Expanding $\phi_\gamma(G)$ up to
second order in $\gamma$ and computing the integral order by order we obtain
\begin{equation}
\langle G\rangle_\gamma\approx \frac{N_2}{1+N_2}-\frac{N_2}{2+N_2}\gamma-\frac{N_2}{(2+N_2)(3+N_2)}\gamma^2
\end{equation}
in full agreement with \eqref{Gaverage} with $N_1=1$.

\begin{figure}[!t]
\includegraphics[scale=0.6,clip]{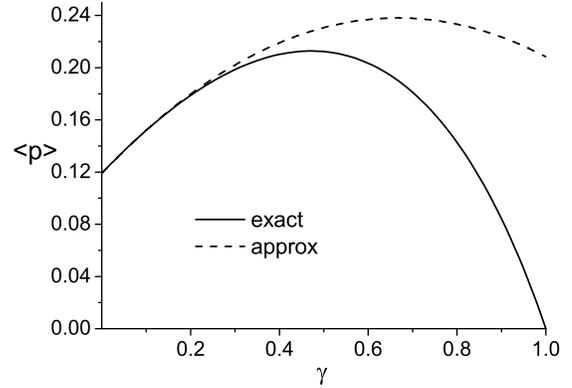}
\caption{Average shot-noise as function of $\gamma$ for $N_1=1$ and $N_2=5$. Solid and
dashed lines are, respectively, the exact result and our approximation. }
\end{figure}

In Figure 1 we plot the average and variance of conductance as functions of $\gamma$ when
$N_1=$ and $N_2=5$, comparing the exact integration of formula \eqref{formulaG} and our
approximate expansions \eqref{Gaverage} and \eqref{varapprox}. The approximation is
excellent for the average, while for the variance the quality deteriorates for $\gamma$
close to $1$.

\section{Average shot-noise up to second order}

Another important quantity that can be measured in the transport context is
shot-noise.\cite{shot} This is related to fluctuations of the electric current as a time
series. Since it is evaluated at zero temperature, it is of quantum nature, arising from
the granularity of electric charge. In terms of reflection eigenvalues, shot noise is
given by \be p(R)=\sum_{i=1}^{N_1}R_i(1-R_i)=s_1(R)-s_2(R)+s_{11}(R).\ee

Its average value is known in the ideal case.\cite{sommers_savin_06} Using the present
approach, we include $\gamma$-effects up to second order. The result is
\begin{align}\label{shot} \langle p \rangle_\gamma\approx
\frac{N_1^2N_2^2}{N(N^2-1)}+\frac{N_2^2(N_1-N_2)^2{\rm Tr}\gamma}{(N^2-1)(N^2-4)}+\nonumber\\
\frac{N_2^2[A_2
(\mathrm{Tr}\gamma)^2-B_2\mathrm{Tr}\gamma^2]}{(N^2-1)(N^2-4)(N^2-9)},\end{align} where
\be A_2=N(5N_2^2-4N_1N_2+N_1^2)+N_1-5N_2,\ee and \be
B_2=N^2N_2^2+4N_2^2-14N_1N_2+3N_1^2+3.\ee The limit of large numbers of channels, with
$N_1=N_2$, can easily be obtained as \be \lim_{N_1=N_2\to\infty}\frac{\la
p\ra_\gamma}{N_1} \approx\frac{1}{8}+\frac{({\rm tr}\gamma)^2-{\rm tr}\gamma^2}{16}.\ee

We compare the approximation (\ref{shot}) against the exact result for $N_1=1$ and
$N_2=5$ in Figure 2 (the exact result is obtained by numerical integration of $G(1-G)$
times the density (\ref{formulaG})). The approximation is not able to account for the
fact that the noise vanishes at $\gamma=1$ (since all particles are surely reflected),
but it can be very good for moderate $\gamma$.

\section{Conclusion}

In summary, combining the theory of symmetric functions and generalized Selberg integrals
we presented a systematic perturbation theory in the opacity matrix $\gamma$ for the jpd
of reflection eigenvalues in chaotic cavities with $\beta=2$ and supporting one ideal and
one non-ideal leads. This jpd is found to be given by the standard Jacobi ensemble
\eqref{jpdideal}, valid for the ideal case, times a correction that can be systematically
expanded in $\gamma$ (see \eqref{jpdnonideal}). Using this result, we computed the
average and variance of conductance, as well as average shot-noise, up to the second
order in $\gamma$ and moments of conductance to leading order.

Our results are valid for arbitrary $N_1,N_2$, in contrast with previously available
results which are exact in $\gamma$ but perturbative in $N_1,N_2$ and often limited to
the leading order term as $N\to\infty$. Comparison with numerics for $N_1=1$ showed that
our perturbative expressions are generally rather accurate for moderate $\gamma$, and
have the advantage of a complete analytical tractability.

Naturally, it would be interesting to extend this calculation to higher orders in
$\gamma$. However, the expressions become quite cumbersome. This may be related to the
asymmetric role of the parameters $N_1$ and $N_2$. Therefore, it would be even more
desirable to be able to consider both leads as non-ideal. Extensions to other symmetry
classes is another challenging open problem.

This work has been partly supported by grant 2011/07362-0, S\~ao Paulo Research
Foundation (FAPESP) and by project Labex PALM-RANDMAT.

\end{document}